\documentclass[twocolumn,showpacs,preprintnumbers,amsmath,amssymb]{revtex4}
\usepackage[dvips]{graphicx} 
\usepackage{dcolumn} 
\newcommand{\beq}{\begin{equation}} 
\newcommand{\beqar}{\begin{eqnarray}} 
\newcommand{\eeq}[1]{\label{#1} \end{equation}} 
\newcommand{\eeqar}[1]{\label{#1} \end{eqnarray}} 
 
\begin{document} 
%%%%%  Title  %%%%% 
\title{Five-Dimensional Fission-Barrier Calculations from 
$^{70}$Se to $^{252}$Cf} 
%%%%%  Authors, affiliations  %%%%%% 
\author{ Peter M\"{O}LLER$^{1}$}\email{moller@moller.lanl.gov}
 \author{Arnold J. SIERK$^{1}$} 
\author{Akira IWAMOTO$^{2}$ }
  \affiliation{$^{1}$Theoretical Division, Los Alamos National Laboratory, Los 
Alamos, New Mexico 87545, USA \\
$^{2}$Japan Atomic Energy Research 
Institute, Tokai-mura, Naka-gun, Ibaraki, 319-1195 Japan} 

\date{\today}

%%%%%  Abstract  %%%%% 
\begin{abstract}

We present fission-barrier-height calculations for nuclei throughout the
Periodic Table based on a realistic macroscopic-microscopic model.
Compared to other calculations: (1) we use a deformation space of
a sufficiently high dimension, sampled densely enough to describe the 
relevant topography of the fission potential,
(2) we unambiguously find the physically relevant saddle points in this space, 
and (3) we formulate our model so that we obtain continuity of the potential energy 
dat the division point between a single system and separated fission fragments or 
colliding nuclei, allowing us to (4) describe both fission-barrier heights 
and ground-state masses throughout the Periodic Table. 

%%%%%  Keywords  %%%%% 
 
\end{abstract}

 
\maketitle

%%%%%  Text  %%%%% 

It has been notoriously difficult to  calculate in a consistent
theoretical model with microscopic content both fission barriers and
ground-state masses for nuclei throughout the Periodic Table.  So far this has
only been possible in the framework of a macroscopic-microscopic
model~\cite{moller81:b,moller95:b}.  However, developments in the area of
nuclear fission show that these earlier studies can now be improved.

On the experimental side it became clear that the barrier data for the two
lightest nuclei previously used in the determination of model constants are
incorrect. In addition, a series of measurements of
fission-barrier heights of nuclei with atomic numbers in the neighborhood of 
$A = 100 $ are now available~\cite{mcmahan85:a,delis91:a,jing99:a}.  Barrier 
heights calculated for these light nuclei using the model from 
Ref.~\cite{moller81:b} are from 1 to 5 MeV too low~\cite{sierk85:a,sierk86:b}.

On the theoretical side we have recently shown that five-dimensional deformation
spaces with the potential energy defined on millions of points are necessary to
determine properly the details of the potential energy such as the locations and
heights of the fission saddle points~\cite{moller00:a,moller01:a}.  This large
deformation space is in stark contrast to one with three degrees of freedom and 
175 deformation points previously used to determine the locations and heights of
saddle points~\cite{moller81:b,moller95:b}.  Moreover, it has been clear for some
time that the Wigner and $A^0$ macroscopic terms in these nuclear mass models
must have a shape dependence~\cite{myers77:a}. This was ignored in previous 
global calculations~\cite{moller81:b,moller95:b,sierk85:a, sierk86:b} and only
incorporated in some of our more limited fission-barrier studies, for example
\cite{moller89:a}.

In nuclear fission the nucleus evolves from a single ground-state shape into two
separated fission fragments. During the shape and configuration changes that
occur in this process the total energy of the system initially increases up
to a maximum, the fission-barrier height, then decreases. 
Calculations of fission barriers involve the determination of the total nuclear
potential energy for different nuclear shapes. Such a calculation defines an
energy landscape as a function of a number of shape coordinates. The 
fission-barrier height is given by the energy relative to the ground state
of the most favorable saddle point that
has to be traversed when the shape evolves from a single shape to separated
fragments. 
We use a technique borrowed from the
area of geographical topography studies, namely immersion (``imaginary water flow'')
\cite{luc91:a,mamdouh98:a,hayes00:a}, to determine the structure of the 
high-dimensional fission potential-energy surfaces~\cite{moller00:a,moller01:a}.

%%%%%  Figure   %%%%% 
\begin{figure}[t] 
 \begin{center} 
 \includegraphics[width=8.0cm,clip]{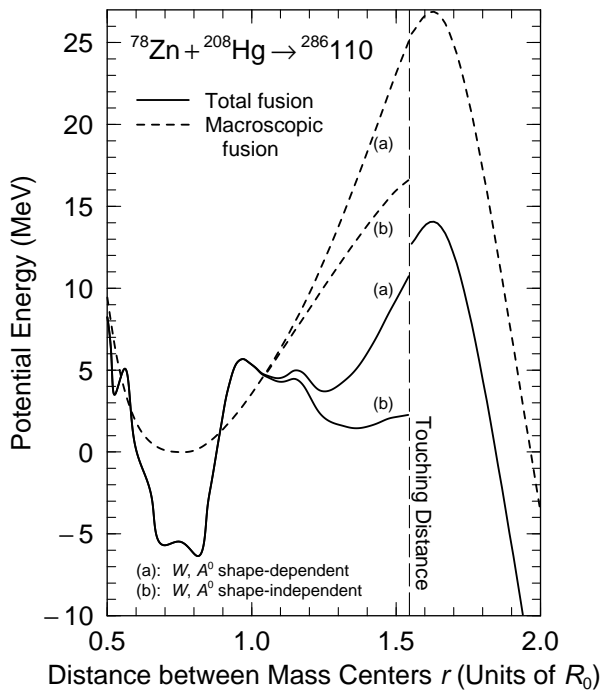} 
   \caption{Calculated macroscopic and total potential energies,   for 
shape sequences leading to the touching configuration, at the 
long-dashed line, of spherical $^{78}$Zn and $^{208}$Hg.
To the left the calculations trace the energy for a {\it single, joined} 
shape configuration  from oblate shapes 
through the spherical shape at $r= 0.75$ to the touching configuration 
at $r=1.52$; to the right the calculations trace the energy for separated 
spherical nuclei beyond the touching point. 
The continuous path through five-dimensional space from the ground state to the touching
configuration is arbitrary; the key point is that the limiting shapes when 
approaching the line of touching from the left and right are identical,
namely spherical $^{78}$Zn and $^{208}$Hg in contact. At a specific value 
of $r$ all curves are calculated for the same shape. 
To obtain continuity of the macroscopic energy at touching, a crucial 
feature in realistic models, it is essential that various model terms depend  
appropriately on nuclear shape, as is the case for the curves (a). The 
slight remaining discontinuity in the {\it total} fusion energy curve arises because 
the Fermi surfaces of the nuclei readjust at touching, and because pairing and 
spin-orbit strength parameters also undergo small discontinuous changes there. 
} 
\label{wigplot} 
 \end{center} 
\end{figure}

A number of models exist for the nuclear potential energy.  At first sight, it
would seem attractive to employ a self-consistent mean-field (SMF) model using
effective forces, for example a Hartree-Fock (HF) or Hartree-Fock-Bogolyubov
(HFB) model with Skyrme or Gogny effective interactions~\cite{aberg90:a,egido97:a,berger89:a}, or a
relativistic Dirac-Hartree model with scalar and vector interactions~\cite{nikolaus92:a}. 
Global HF mass calculations have recently been presented~\cite{goriely01:a,goriely02:a}.  However, at
least two major problems with such calculations remain unresolved. First, no
effective force has been found which can describe both nuclear masses and
fission barriers for nuclei of all mass numbers. 
Second, even if
an appropriate effective interaction could be determined, it is extremely
difficult, and in practice has so far proven impossible unambiguously to locate
an actual saddle-point configuration in  SMF models. There exists a common
misconception that constrained self-consistent HF or HFB calculations with
Skyrme or Gogny forces automatically take into account all non-constrained shape
variables in a proper manner. In fact, the apparent saddle points that
appear in constrained HF calculations in the general case have no relation to
the true saddle points; we give an illustrative example below.

For calculating the fission potential-energy surface in this work, we adopt 
 the macroscopic-microscopic finite-range liquid-drop model
(FRLDM)~\cite{moller95:b} generalized to account for
all required shape-dependencies of its various macroscopic terms. 
In contrast the finite-range droplet model (FRDM)~\cite{moller95:b}
cannot be generalized in this way. In Fig.~1
we illustrate the large discontinuities that occur at the transition point
between single and separated systems when the shape dependences of
the Wigner and $A^0$ macroscopic energies are neglected, and
the continuous behavior exhibited for the current formulation.

In the macroscopic-microscopic approach,  it is important that the shape
description be flexible enough to allow accessing those configurations which are
physically important to the fission process.  In addition to the commonly used
elongation, necking and mass-asymmetry degrees of freedom, it is essential to
include the deformations of the partially formed fragments.  This is
because the microscopic binding due to fragment shell structure, which is
sensitive to the fragment deformations, can be as large as 5 MeV even for shapes with
a fairly large neck radius. We use the three-quadratic-surface (3QS) shape
parameterization~\cite{moller01:a,nix69:a} to describe shapes with these five degrees 
of freedom. By investigating the scale over which the microscopic energy varies
significantly, we determine the coordinate mesh upon which we need to define the
energy.  We find that we get a reasonable coverage of the space by
defining a grid of 15 points each in the neck diameter and left and right
fragment deformations, 20 points in the mass asymmetry, and 41 points in the
nuclear elongation. A few grid points 
do not refer to real shapes, so we
are left with a 
grid of 2,610,885 points~\cite{moller01:a}.  

In an unconstrained mean-field calculation, one starts with some initial
density, usually defined in terms of trial wavefunctions, then determines new
wavefunctions which are solutions of the potential derived from the density. By
iterating to convergence one finds a local minimum: the nuclear
ground state or possibly a fission-isomeric state. To try to find a fission 
barrier, some have chosen to solve a constrained problem, which leads to the
minimum-energy state subject to the constraint, often taken to be the
quadrupole moment.  By applying a series of constraints with increasing
deformation, a curve of energy as a function of constrained deformation is
found. Such curves often exhibit discontinuities and may not pass
through the real saddle point in multidimensional space as is
discussed in more detail in Refs.~\cite{moller00:a,myers96:a}.

In a macroscopic-microscopic calculation one should in principle be able to
locate saddle points by solving for all shapes that have a zero derivative of
the energy with respect to all the degrees of freedom. 
This method works for a
purely macroscopic model~\cite{sierk86:b}, 
but macroscopic-microscopic models using the Strutinsky shell-correction 
technique~\cite{strutinsky67:a,strutinsky68:a}, are subject to fluctuations 
when small shape changes are made, making it
difficult to obtain accurate derivatives by numerical techniques.
Even if all saddle points in a high-dimensional space could be found in
this way, one must still understand the topography and deduce which
saddle would correspond to the actual peak of the barrier. 
Before the breakthrough study in Ref.~\cite{mamdouh98:a}, what has usually
been done in calculations involving more than two degrees of freedom is 
to first define a
two-dimensional space of two primary shape coordinates. For each point in this
two-dimensional space the energy is then minimized with respect to a set of 
additional shape degrees of freedom.  It was incorrectly
assumed that if no discontinuities occurred in this two-dimensional
surface then its saddle points would be identical to the saddle-points in the full,
higher-dimensional space.  For all but the most structureless functions
this procedure is incorrect and may actually result in more inaccurate
saddle points than if only the original two-dimensional space is studied.
%%%%%  Figure   %%%%% 
 \begin{figure}[t]
\begin{center}
\includegraphics[width=8.0cm,clip]{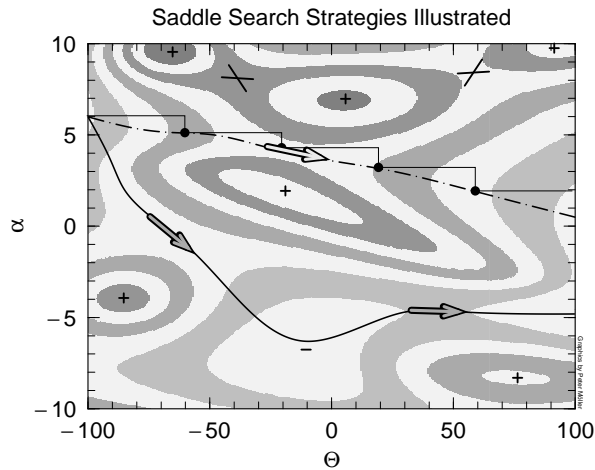} 
\caption{
Maxima ($+$), minima ($-$), and 
saddle points (arrows or crossed lines) of a two-dimensional function. As discussed 
in the text it is not possible to obtain a lower-dimensional representation of
this surface by ``minimizing'' with respect to the ``additional'' ($\alpha$)
shape degree of freedom. Darker shades of gray indicate higher function
values. Alternate contour bands are light gray for readability.}
\label{minfalil2g} 
\end{center}
\end{figure}

We illustrate in Fig.~2 some of the problems that one may 
encounter in either
a constrained SMF calculation or in a macroscopic-microscopic model
when attempting to reduce the dimensionality of a problem via minimization
while preserving the essential features of the potential energy. We 
assume the $\Theta$ coordinate corresponds to the fission direction and
$\alpha$  to all other degrees of freedom. Because of the
multiple saddle points, it is not clear {\it a priori} which one would
correspond to the fission threshold.  We identify the point $\Theta = -100$,
$\alpha = 6$ as the ground state or fission-isomer state and proceed to find a 
``constrained'' fission barrier.  From the starting point we increase $\Theta$ 
by 40 (smaller steps will not alter the result) while keeping $\alpha$
fixed. From the new position we then minimize with respect to $\alpha$ and
find ourselves at the first black dot. When we repeat this process we obtain
the dot-dashed curve. The energy along this path is a continuous function
and  the white arrow would be identified as the fission saddle point.
However, this saddle is higher than those shown by gray arrows, 
which can only be identified when the full space is explored.
Of course in a constrained SMF calculation the convergence towards
a solution is more complex than ``sliding downhill'', since the wave
functions and potential change during this process. 
However, solutions of constrained SMF equations do show similar behavior;
often converging to a local minimum which depends on the starting configuration,
a process similar to what is sketched in Fig.~2.

The fundamental point is that the fission saddle point can only be determined
%%%%% Table 1  %%%%% 
\begin{table}[b!] 
{\small 
\fontsize{9}{12} 
 \begin{center} 
  \caption{ Macroscopic model parameters of the FRLDM (1992) and those
 obtained in the present adjustment, designated FRLDM (2002) using barrier heights obtained 
in our five-dimensional calculation.} 
   \label{macconst} 
   \begin{tabular}{rrr} 
\hline    
  Constant & FRLDM (1992) & 
FRLDM (2002)  \\ 
\hline 
\vspace{-2mm} \\ 
$  a_{\rm v}\phantom{0!}$         & 16.00126 MeV & 16.02500 MeV \\ 
$  \kappa_{\rm v}\phantom{0!}$    &  1.92240 MeV &  1.93200 MeV \\ 
$  a_{\rm s}\phantom{0!}$         & 21.18466 MeV & 21.33000 MeV \\ 
$  \kappa_{\rm s}\phantom{0!}$    &  2.34500 MeV &  2.37800 MeV \\ 
$  a_0\phantom{0!}$               &  2.61500 MeV &  2.04000 MeV \\ 
$  c_{\rm a}\phantom{0!}$         &  0.10289 MeV &  0.09700 MeV \\     
\hline 
  \end{tabular} 
 \end{center}
}
\end{table}
from global properties of the multidimensional energy surface, not
from local excursions from a specific starting point.  
We therefore implement the immersion method mentioned above and
first identify
all minima by locating the points which have a lower energy than all $3^n - 1$
neighboring points in n-dimensional coordinate space.  We then progressively
fill up the ground-state minimum with ``water'', determining when a prespecified
 point in the fission valley becomes ``wet''. By adjusting the increase in
the water level carefully we are able unambiguously to identify the location and
energy of the grid point nearest to the true  saddle. 

%%%%%  Figure   %%%%% 
\begin{figure}[t] 
 \begin{center} 
 \includegraphics[width=7.5cm,clip]{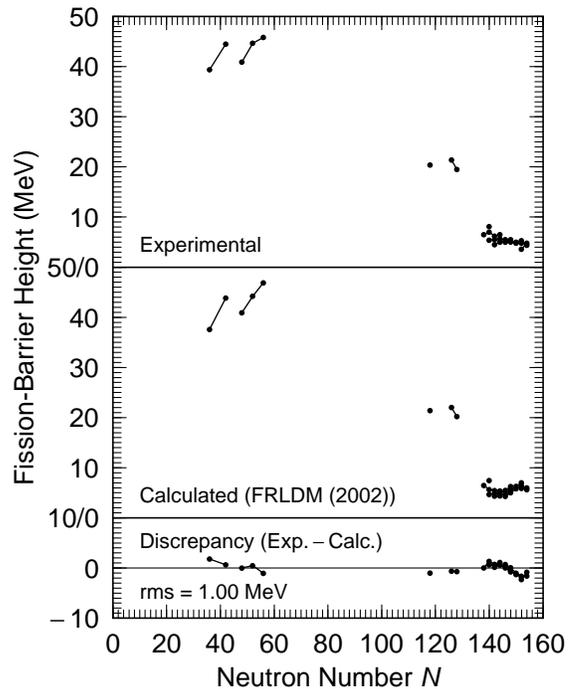} 
   \caption{Comparison of calculated and experimental fission-barrier 
heights for nuclei throughout the Periodic Table, after a  
readjustment of the macroscopic model constants. 
Experimental barriers are 
well reproduced by the calculations, the rms error is only 0.999 MeV for 
31 nuclei. In the actinide region it is the outer of the  
two peaks in the ``double-humped'' barrier 
that is compared to experimental data. } 
   \label{plobar2002} 
 \end{center} 
\end{figure} 

Because we are studying a higher dimensional deformation space with 
over 10000 times as many points as in
previous calculations, we find that the saddles for a
given nucleus are always lower than those found for the same model 
parameters in our earlier studies~\cite{moller95:b,moller00:a}. 
This means we need to redetermine the
parameters of our nuclear-structure model. Because new
parameter sets may change the location of
the saddle points and minima, an iterative procedure is in principle
required. However these changes in deformation are small; here we
do just the first iteration and vary six parameters
of the macroscopic energy functional to optimize the reproduction of both
ground-state binding energies and fission-barrier heights. This process is
identical to the one  followed in the creation of our 1995 mass 
table~\cite{moller95:b}. We show in Table~\ref{macconst} the constants from 
Ref.~\cite{moller95:b} (determined in 1992) and
those obtained in our readjustment taking into account the larger
deformation space in locating the fission saddle points.  We
have made a number of tests which allow us to conclude that a self-consistent
redetermination of the ground-state and saddle-point deformations would change
our calculated energies by less than 0.1 MeV. 

These constants (in particular $a_0$) differ slightly from the preliminary
%%%%% Table 2  %%%%% 
\begin{table}[t] 
{\small 
\fontsize{9}{12} 
 \begin{center} 
   \caption{ Calculated barriers for 31 nuclei compared to experimental
data. The first 5 barriers are macroscopic barriers.} 
   \label{barrier} 
   \begin{tabular}{rrrrrrrrrrr} 
\hline    
   \multicolumn{1}{c}{$Z$}              &   
   \multicolumn{1}{c}{$A$}              &
   \multicolumn{1}{c}{$E_{\rm exp}$}    & 
   \multicolumn{1}{c}{$E_{\rm calc}$}   &
   \multicolumn{1}{c}{$\Delta E$}       &
                                        &
   \multicolumn{1}{c}{$Z$}              &   
   \multicolumn{1}{c}{$A$}              &
   \multicolumn{1}{c}{$E_{\rm exp}$}    & 
   \multicolumn{1}{c}{$E_{\rm calc}$}   &
   \multicolumn{1}{c}{$\Delta E$}          \\
                                        &        
                                        &   
\multicolumn{1}{c}{(MeV)}               &  
\multicolumn{1}{c}{(MeV)}               &   
\multicolumn{1}{c}{(MeV)}               &
                                        &
                                        &        
                                        &   
\multicolumn{1}{c}{(MeV)}               &  
\multicolumn{1}{c}{(MeV)}               &   
\multicolumn{1}{c}{(MeV)}                   \\
\hline 
\vspace{-2mm} \\ 
   34     &  70    &   39.40     &  37.58   &     1.81  &&    92 & 238 & 5.50 &   5.48     &   0.01     \\
   34     &  76    &   44.50     &  43.84   &     0.65  &&    92 & 240 & 5.50 &   6.27     &  $-0.77$     \\
   42     &  90    &   40.92     &  40.92   &    $-0.00$  &&    94 & 236 & 4.50 &   4.35     &   0.14     \\
   42     &  94    &   44.68     &  44.20   &     0.47  &&    94 & 238 & 5.00 &   4.39     &   0.60     \\
   42     &  98    &   45.84     &  46.88   &    $-1.04$  &&    94 & 240 & 5.15 &   4.83     &   0.31     \\ 
\cline{1-5}						  
   80     & 198    &   20.40     &  21.41   &    $-1.01$  &&    94 & 242 & 5.05 &   5.55     &  $-0.50$     \\
   84     & 210    &   21.40     &  22.02   &    $-0.62$  &&    94 & 244 & 5.00 &   6.29     &  $-1.29$     \\
   84     & 212    &   19.50     &  20.20   &    $-0.70$  &&    94 & 246 & 5.30 &   7.01     &  $-1.71$     \\
   88     & 228    &    8.10     &   7.45   &     0.64  &&    96 & 242 & 5.00 &   4.28     &   0.71     \\
   90     & 228    &    6.50     &   6.47   &     0.02  &&    96 & 244 & 5.10 &   5.02     &   0.07     \\
   90     & 230    &    7.00     &   5.65   &     1.34  &&    96 & 246 & 4.80 &   5.81     &  $-1.01$     \\
   90     & 232    &    6.20     &   5.45   &     0.74  &&    96 & 248 & 4.80 &   6.41     &  $-1.61$     \\
   90     & 234    &    6.50     &   5.36   &     1.13  &&    96 & 250 & 4.40 &   5.98     &  $-1.58$     \\
   92     & 232    &    5.40     &   4.67   &     0.72  &&    98 & 250 & 3.60 &   5.88     &  $-2.28$     \\
   92     & 234    &    5.50     &   4.89   &     0.60  &&    98 & 252 & 4.80 &   5.63     &  $-0.83$     \\
   92     & 236    &    5.67     &   4.98   &     0.68  &&       &     &      &            &            \\
   \hline 
  \end{tabular} 
 \end{center}
}
\end{table}
set presented in Ref.~\cite{moller02:b}. Those constants were affected by an
error in the expression for the $a_0$ energy term in the constant-adjustment 
program. However, none of the other previous results, conclusions or figures 
were
affected significantly by this computer-program bug. Here we have checked the
calculation of macroscopic-model saddle-point shapes and energies by the use 
of two independently written codes. The microscopic energy
model is unchanged from~\cite{moller95:b}.

The 1992 calculation reproduced an experimental 1989 nuclear mass table 
\cite{audi89:a} with a model error of 0.779 MeV, and 28 barrier heights
with model error of 1.4 MeV.  The revised data set 
here \cite{delis91:a,jing99:a,madland01:a} incorporates seven new experimental 
barrier heights  and
removes four old ones.  The fit to the revised table of 31 barriers  has 
an rms error of 0.999 MeV, and the fit to the  same 1989 mass table has a model error of 0.752 MeV
using the parameters in the last column of Table I. 
We show the experimental and
calculated barrier heights as well as remaining discrepancies in Table II and Fig.~3. 

This work was supported by the U. S. DOE.

%%%%%  References  %%%%% 

%\bibliographystyle{unsrt}

\end{document}